\documentclass{INTERSPEECH2023}

\usepackage{inconsolata}
\usepackage{booktabs} 
\usepackage{pifont}
\usepackage{graphicx} 
\usepackage{multirow}
\usepackage{multicol}
\usepackage{arydshln}
\usepackage{color}
\usepackage{amssymb}
\usepackage{cancel}
\usepackage{diagbox}
\usepackage{makecell}
\usepackage{xcolor}
\usepackage[T1]{fontenc}
\interspeechcameraready 

\usepackage{hyperref}
\hypersetup{
    colorlinks=true
    }
    
\newcommand\blfootnote[1]{%
  \begingroup
  \renewcommand\thefootnote{}\footnote{#1}%
  \addtocounter{footnote}{-1}%
  \endgroup
}

\title{Audio-CoT: Exploring Chain-of-Thought Reasoning \\ in Large Audio Language Model}
\name{Ziyang Ma$^{1,2}$, Zhuo Chen$^3$, Yuping Wang$^3$, Eng Siong Chng$^2$, Xie Chen$^1$$^\dag$}
\address{
  $^1$MoE Key Lab of Artificial Intelligence, X-LANCE Lab, Shanghai Jiao Tong University, \\
  $^2$Nanyang Technological University, \\
  $^3$ByteDance Inc. 
  }
\email{\{zym.22, chenxie95\}@sjtu.edu.cn}

\begin{document}
\maketitle
 
\begin{abstract}
\blfootnote{Corresponding author$^\dag$.}
Large Audio-Language Models (LALMs) have demonstrated remarkable performance in tasks involving audio perception and understanding, such as speech recognition and audio captioning. 
However, their reasoning capabilities—critical for solving complex real-world problems—remain underexplored. 
In this work, we conduct the first exploration into integrating Chain-of-Thought (CoT) reasoning into LALMs to enhance their reasoning ability across auditory modalities. 
We evaluate representative CoT methods, analyzing their performance in both information extraction and reasoning tasks across sound, music, and speech domains. 
Our findings reveal that CoT methods significantly improve performance on easy and medium tasks but encounter challenges with hard tasks, where reasoning chains can confuse the model rather than improve accuracy. Additionally, we identify a positive correlation between reasoning path length and accuracy, demonstrating the potential of scaling inference for advanced instruction-following and reasoning. This study not only highlights the promise of CoT in enhancing LALM reasoning capabilities but also identifies key limitations and provides actionable directions for future research.

\end{abstract}

\section{Introduction}
Humans possess an extraordinary ability to perceive, understand, and reason about a wide variety of auditory signals, ranging from speech and music to general environmental sounds. 
This capability enables us to draw inferences, solve problems, and adapt to complex auditory scenarios. 
With the rapid progress of large language models (LLMs) and significant breakthroughs in the audio domain, large audio-language models (LALMs)~\cite{tang2023salmonn, gong2023joint, chu2023qwen, Qwen2-Audio, ghosh2024gama, kong2024audio} have been developed to process and analyze multi-modal audio data. 
These models, equipped with advanced architectures and large-scale training, have shown remarkable performance in tasks related to audio perception and understanding, such as automatic speech recognition (ASR)~\cite{li2023prompting, wu2023decoder, ma2024embarrassingly, fathullah2023prompting, yu2023connecting, yang2024mala, yang2024ctc, geng2024unveiling, bai2024seed} and automated audio captioning (AAC)~\cite{wu2023beats, chen2024slam, li2024drcap, liu2024leveraging}. 
However, the reasoning capabilities of LALMs—critical for bridging the gap between human-like cognition and artificial intelligence—remain largely underexplored. 

\begin{figure}[t]
  \centering
  \includegraphics[width=0.8\linewidth]{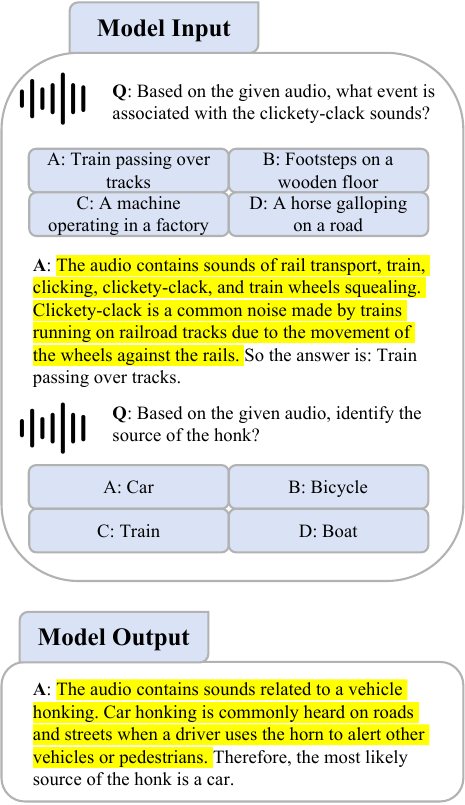}
  \caption{Illustration of chain-of-thought (CoT) reasoning in large audio language models. The Manual-CoT approach with few-shot learning is taken as an example to demonstrate its mechanism. Reasoning thoughts are highlighted.}
  \label{fig:cot}
\end{figure}

Existing research on LALMs primarily focuses on sensory-level tasks, aiming to transcribe, classify, or summarize audio content~\cite{huang2024dynamic, yang2024air, wang2024audiobench, shi2024versa}. 
While these achievements are significant, they fall short of addressing the higher-level reasoning required for more complex tasks, such as multi-step inferences or logical deduction from auditory inputs. 
For instance, reasoning about the intent behind a spoken statement, the emotional context of a piece of music, or the cause-and-effect relationships in environmental sounds requires structured cognitive processes beyond simple understanding. 

To address this gap, we explore Chain-of-Thought (CoT)~\cite{wei2022chain} reasoning within the context of LALMs. 
CoT reasoning, a technique initially introduced for LLMs, facilitates step-by-step logical deduction by prompting the model to generate explicit reasoning paths. 
As illustrated in Figure~\ref{fig:cot}, CoT reasoning leverages structured thought processes to enable models to make more accurate and interpretable predictions. 
Using the Manual-CoT approach with few-shot learning as an example, the figure demonstrates how explicit reasoning steps can be employed to guide the model toward better results. 

In this work, we systematically evaluate representative CoT methods including Manual-CoT, Zero-Shot-CoT, and Desp-CoT. 
Our findings reveal that CoT methods significantly enhance performance on easy and medium tasks but face challenges on hard tasks, where reasoning chains sometimes confuse the model instead of clarifying complex problems. Furthermore, the application of Self-Consistency further improves performance by marginalizing multiple reasoning paths. 
Additionally, we observe a positive correlation between reasoning length and performance, highlighting the potential of scaling inference processes. 
This study represents the first exploration of CoT reasoning applied to LALMs, emphasizing the importance of integrating reasoning capabilities into multi-modal audio understanding systems.

\section{Related Work}
\subsection{Large Audio Language Model}
With the rapid development of LLMs and huge advancements in the audio domain, LALMs have garnered increasing attention for their powerful general audio understanding capabilities~\cite{latif2023sparks, wu2024towards, peng2024survey}. 
Similar to multi-modal large language models (MLLMs), LALMs typically consist of three core components: an audio encoder for modality-specific perception, an LLM for text generation, and a projector that bridges the encoder and the language model. 
Some work has focused on improving LALM performance through innovations in model architecture. 
For example, SALMONN~\cite{tang2023salmonn} employs dual encoders to separately process speech and audio signals, effectively mitigating domain conflicts between different audio sub-modalities. 
Other research efforts have explored training strategies to enhance LALMs. A notable example is Qwen2-Audio~\cite{Qwen2-Audio}, which follows a comprehensive LLM training pipeline comprising pre-training, supervised fine-tuning (SFT), and reinforcement learning with human feedback (RLHF).
Moreover, expanding the scale and diversity of instruction-tuning data~\cite{ghosh2024gama} has further enabled LALMs to proficiently handle diverse modalities, including speech, music, and general sound. 
Some models excel in jointly processing two modalities~\cite{gong2023joint, ghosh2024gama}, while others have successfully extended this capability to three modalities simultaneously~\cite{tang2023salmonn, chu2023qwen, Qwen2-Audio}.

However, most existing LALM research remains focused on perception and understanding tasks, with relatively little attention given to reasoning capabilities. 
Developing reasoning abilities in LALMs is a critical step toward achieving advanced audio understanding, and represents a significant milestone on the path to artificial general intelligence (AGI). 

\subsection{Chain-of-Thought Reasoning}
For complex problems, LLMs often struggle to directly provide correct answers. The CoT approach addresses this challenge by prompting models to explicitly output intermediate reasoning steps before arriving at the final answer. 
This method has been shown to significantly enhance the arithmetic, commonsense, and logical reasoning capabilities of LLMs~\cite{chu2024navigate}. 
CoT techniques can generally be categorized into manually constructed~\cite{wei2022chain} and automatically generated~\cite{kojima2022large, zhang2022automatic} approaches. 
Manually constructed CoT~\cite{wei2022chain} relies on carefully designed prompt examples, which typically result in high-quality reasoning chains but come with substantial human labor costs. 
On the other hand, automatically generated CoT~\cite{kojima2022large, zhang2022automatic} leverages specific prompt texts to induce models to generate reasoning chains without examples. While more scalable, the automatic method often suffers from lower quality, leading to severe hallucination issues. 
To mitigate these limitations, some studies have explored hybrid methods that combine manual and automatic construction to achieve better outcomes~\cite{liu2021generated, xu2023expertprompting}.

In the multimodal domain, several works have extended CoT reasoning to multi-modal chain-of-thought (MCoT) reasoning to enhance the performance of vision-language models (VLMs)~\cite{zhang2023multimodal, wu2023role}. 
More recently, efforts such as investigating reasoning complexity~\cite{chen2024m} and post-training scalability~\cite{guo2024mammoth}, provide further insights into improving MCoT techniques. 
However, in the context of LALMs, no exploration of CoT methodologies has been conducted to date, which represents a notable gap in the field of general audio understanding and reasoning. 

\begin{table*}[t]
\centering
\caption{Accuracy (\%$\uparrow$) comparison of the baseline results and various CoT methods on the MMAU benchmark, evaluated across the sound, speech, and music modalities. The ``Original Results'' are derived directly from the MMAU paper, where the answers are extracted using rule-based methods, potentially underestimating its true performance. To address this limitation, we reproduced the results using a more robust methodology by leveraging the GPT-4o-mini API to regularize the answers, ensuring consistency and accuracy in the evaluation. The best-performing results are highlighted in \textbf{bold}, while the second-best results are \underline{underlined}. For reference, we also include results for random guessing, the most frequent choice, and human evaluations as reported in the original MMAU paper. }
\label{tab:main_results}
\resizebox{\linewidth}{!}{
\begin{tabular}{lcccccccccccc}
\toprule \toprule
\multirow{2.5}{*}{\textbf{Methods}} & \multicolumn{4}{c}{\textbf{Information Extraction}} & \multicolumn{4}{c}{\textbf{Reasoning}} & \multicolumn{4}{c}{\textbf{Total}} \\
\cmidrule{2-13}
& \textbf{Sound} & \textbf{Music} & \textbf{Speech} & \textbf{Total} & \textbf{Sound} & \textbf{Music} & \textbf{Speech} & \textbf{Total} & \textbf{Sound} & \textbf{Music} & \textbf{Speech} & \textbf{Total} \\
\midrule \midrule
\multicolumn{1}{l}{\textbf{\textit{Baselines}}} \\ 
Original Results & - & - & - & - &  - & - & - & - & 54.95 & 50.98 & 42.04 & 49.20 \\
Reproductive Results & 76.34 & 54.70 & 10.00 & 58.50 & 43.33 & 45.75 & 45.37 & 44.76 & 52.55 & 50.60 & 43.24 & 48.80 \\
\quad + Answer Normalization & \underline{82.80} & 56.35 & \textbf{40.00} & 63.61 & 52.50 & \textbf{53.59} & 51.44 & 52.27 & 60.96 & 55.09 & 50.75 & 55.60 \\
\midrule \midrule
\multicolumn{1}{l}{\textbf{\textit{CoT Methods}}} \\ 
Manual-CoT & 80.65 & \textbf{60.77} & \textbf{40.00} & \underline{65.65} & \textbf{54.17} & 50.33 & 54.31 & 53.40 & \underline{61.56} & \underline{55.99} & 53.45 & \underline{57.00} \\
Zero-Shot-CoT & \textbf{84.95} & \underline{59.12} & \textbf{40.00} & \textbf{65.99} & \underline{52.92} & \underline{52.94} & \textbf{56.23} & \textbf{54.39} & \textbf{61.86} & \textbf{56.29} & \textbf{55.26} & \textbf{57.80} \\
Desp-CoT & 75.27 & 57.45 & \textbf{40.00} & 61.90 & \underline{52.92} & \textbf{53.59} & \underline{54.95} & \underline{53.97} & 59.16 & 55.69 & \underline{54.05} & 56.30 \\
\midrule \midrule
\multicolumn{1}{l}{\textcolor{gray}{\textbf{\textit{Reference}}}} \\ 
\textcolor{gray}{Random Guess} & \textcolor{gray}{-} & \textcolor{gray}{-} & \textcolor{gray}{-} & \textcolor{gray}{-} & \textcolor{gray}{-} & \textcolor{gray}{-} & \textcolor{gray}{-} & \textcolor{gray}{-} & \textcolor{gray}{26.72} & \textcolor{gray}{24.55} & \textcolor{gray}{26.72} & \textcolor{gray}{26.00} \\
\textcolor{gray}{Most Frequent Choice} & \textcolor{gray}{-} & \textcolor{gray}{-} & \textcolor{gray}{-} & \textcolor{gray}{-} & \textcolor{gray}{-} & \textcolor{gray}{-} & \textcolor{gray}{-} & \textcolor{gray}{-} & \textcolor{gray}{27.02} & \textcolor{gray}{20.35} & \textcolor{gray}{29.12} & \textcolor{gray}{25.50} \\
\textcolor{gray}{Human} & \textcolor{gray}{-} & \textcolor{gray}{-} & \textcolor{gray}{-} & \textcolor{gray}{-} & \textcolor{gray}{-} & \textcolor{gray}{-} & \textcolor{gray}{-} & \textcolor{gray}{-} & \textcolor{gray}{86.31} & \textcolor{gray}{78.22} & \textcolor{gray}{82.17} & \textcolor{gray}{82.23} \\
\bottomrule \bottomrule
\end{tabular}
}
\end{table*}

\section{Methods}
\label{sec:methods}
CoT reasoning, as one of the most significant methods for enhancing reasoning capabilities, has yet to be extensively explored in the context of LALMs. 
In this work, we pioneer the application of CoT methods to LALMs, investigating how they can be adapted to improve performance without altering the model's training paradigm.

\textbf{Manual-CoT}~\cite{wei2022chain} combines the concepts of few-shot learning and chain-of-thought prompting, leveraging the in-context learning capabilities of LLMs to ensure that the model’s output adheres to the reasoning structure provided in handcrafted examples. 
Let the set of examples $\mathcal{E}$ be represented as:
\begin{equation}
    \mathcal{E} = \{(A_i, I_i, C_i)\}_{i=1}^N, 
\end{equation}
where $A_i$ is the $i$-th example audio, $I_i$ is the corresponding instruction, and $C_i$ is the reasoning chain for that example.
For a given input audio $A_{\text{input}}$ and an input instruction $I_{\text{input}}$, the model generates an output reasoning chain $C_{\text{output}}$ by utilizing the example set $\mathcal{E}$:
\begin{equation}
    C_{\text{output}} = f_{\text{LALM}}(\mathcal{E}, A_{\text{input}}, I_{\text{input}}), 
\end{equation}
where $f_{\text{LALM}}$ represents the inference process of the LALM. By combining the examples with CoT reasoning, the model emulates the structured reasoning process delineated in the examples, effectively transferring the manually designed reasoning chain patterns to new inputs.

\textbf{Zero-Shot-CoT}~\cite{kojima2022large} eliminates the need for explicit handcrafted examples and instead relies on a simple natural language magic prompt, such as ``Let's think step by step'' to guide the model in generating reasoning chains. 
By leveraging the in-context reasoning ability of LLMs, Zero-Shot-CoT enables logical step-by-step deduction even without task-specific examples. 
Let the magic prompt be represented as $P$, the model $f_{\text{LALM}}$ enerates an output reasoning chain 
output $C_{\text{output}}$ directly from the augmented instruction, written as:
\begin{equation}
    C_{\text{output}} = f_{\text{LALM}}(A_{\text{input}}, I_{\text{input}} + P). 
\end{equation}
This approach is efficient, scalable, and eliminates the need for curated examples while still achieving high-quality reasoning outputs. 

\textbf{Desp-CoT}~\cite{wu2023role} leverages the strengths of the cross-modal understanding property of LALM to enhance reasoning by generating a descriptive caption of the audio input before initiating the reasoning process. 
This intermediate step provides a natural language representation of the audio content, which serves as an anchor for subsequent reasoning tasks. 
The process is composed of two distinct phases: description generation and reasoning chain generation. 
For the description generation, the input audio $A_{\text{input}}$ is first prompted to produce a descriptive caption $D$, which can be expressed as: 
\begin{equation}
    D = f_{\text{Caption}}(A_{\text{input}}), 
\end{equation}
where $f_{\text{Caption}}$ represents the LALM or other caption models. 
Then the generated caption $D$ is concatenated with the input instruction $I_{\text{input}}$ to form an augmented input instruction, which is used to guide the reasoning chain generation: 
\begin{equation}
    C_{\text{output}} = f_{\text{LALM}}(A_{\text{input}}, D + I_{\text{input}}). 
\end{equation}
This method not only improves the interpretability of the reasoning process but also enhances the performance of complex audio-instruction tasks.

\begin{table}[htbp]
\centering
\caption{Impact of applying Self-Consistency to different CoT methods on Accuracy (\%$\uparrow$) across Sound, Music, Speech, and Total categories on the MMAU benchmark. }
\label{tab:self_consistency}
\resizebox{\linewidth}{!}{
\begin{tabular}{lcccc}
\toprule \toprule
\textbf{Methods} & \textbf{Sound} & \textbf{Music} & \textbf{Speech} & \textbf{Total} \\
\midrule \midrule
Baseline & 60.96 & 55.09 & 50.75 & 55.60 \\
Manual-CoT & 61.56 & 55.99 & 53.45 & 57.00 \\
\quad + Self-Consistency & 61.26 & \textbf{57.78} & 53.45 & 57.50 \\
Zero-shot-CoT & 61.86 & 56.29 & 55.26 & 57.80 \\
\quad + Self-Consistency & \textbf{62.16} & 55.99 & \textbf{56.16} & \textbf{58.10} \\
\bottomrule \bottomrule
\end{tabular}
}
\end{table}

\begin{figure*}[htbp]
  \centering
  \includegraphics[width=\linewidth]{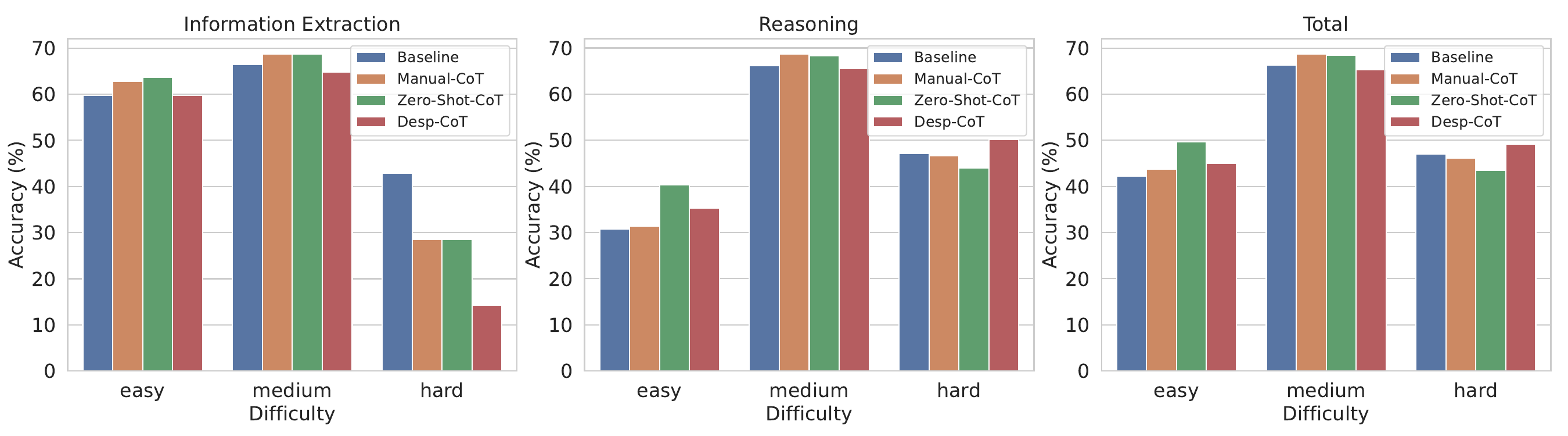}
  \caption{Comparison of different CoT approaches across information extraction, reasoning, and total at three difficulty levels (easy, medium, and hard). }
  \label{fig:difficulty}
\end{figure*}

\section{Experiments}
\subsection{Setup}
For the model setting, Qwen2-Audio-7B-Instruct~\cite{Qwen2-Audio} is adopted. 
Renowned for its strong performance across diverse benchmarks such as AIR-Bench~\cite{yang2024air}, OmniBench~\cite{li2024omnibench}, and MMAU~\cite{sakshi2024mmau}, this model stands out as one of the most powerful open-source LALMs currently available. 
For the dataset setting, we utilize the MMAU benchmark~\cite{sakshi2024mmau}, which is designed to assess LALMs' information extraction and reasoning abilities across three domains: sound, music, and speech. 
Approximately $30\%$ of the questions focus on information extraction tasks, while the remaining $70\%$ are dedicated to reasoning tasks, reflecting a strong emphasis on evaluating logical and inferential capabilities. 
MMAU constructs multiple-choice questions spanning $27$ distinct skills and assigns each question a difficulty level, making it an ideal dataset for investigating CoT reasoning. 
While the full MMAU dataset comprises $10,000$ entries, only a subset of $1,000$ entries has been made publicly available. This subset maintains the same distribution as the full dataset, serving as a reliable evaluation set for CoT reasoning experiments. 
To minimize variability introduced by stochastic sampling, all results are generated with greedy search except for the Self-Consistency method. 

\subsection{Main Results}
Table~\ref{tab:main_results} presents the accuracy comparison of the baseline results and CoT methods mentioned in Section~\ref{sec:methods} on the MMAU benchmark across the sound, music, and speech modalities, considering both information extraction and reasoning tasks. 

Generally speaking, the adoption of CoT methods universally improves performance over the baseline results. Among the CoT methods, Zero-Shot-CoT consistently achieves the best overall performance, with a total accuracy of $57.80\%$, surpassing the baseline with answer normalization ($55.60\%$). Manual-CoT and Desp-CoT also yield significant improvements, achieving total accuracies of $57.00\%$ and $56.30\%$, respectively. These findings demonstrate that CoT-based reasoning effectively enhances the LALM's ability to handle both information extraction and reasoning tasks. 
Besides, we found that the results for reasoning tasks are consistently lower than those for information extraction tasks. 
For example, Manual-CoT achieves $53.40\%$ in reasoning compared to $65.65\%$ in information extraction, while Zero-Shot-CoT achieves $54.39\%$ in reasoning compared to $65.99\%$ in information extraction. 
These results highlight the increased complexity of reasoning tasks, where the model struggles to maintain the same level of performance as in relatively straightforward information extraction tasks. 
Despite the improvements brought by CoT methods, there remains a significant gap between human performance and model performance. 

We further apply Self-Consistency~\cite{wang2022self} on the MMAU benchmark to improve performance. 
Instead of relying solely on a single greedy reasoning path, Self-Consistency first generates a diverse set of reasoning paths through sampling and then determines the most consistent answer by marginalizing the sampled reasoning paths.
In our experiments, we implement it through $5$ sampling iterations followed by majority voting based on maximum probability. 
Table~\ref{tab:self_consistency} illustrates the impact of applying Self-Consistency on accuracy. 
Overall, these results demonstrate that Self-Consistency is a valuable strategy for boosting the performance of CoT methods, particularly when combined with Zero-Shot-CoT, where it achieves the highest total accuracy of $58.10\%$. But its effectiveness may depend on the specific task and modality.

\subsection{Analysis}

\begin{figure}[htbp]
  \centering
  \includegraphics[width=\linewidth]{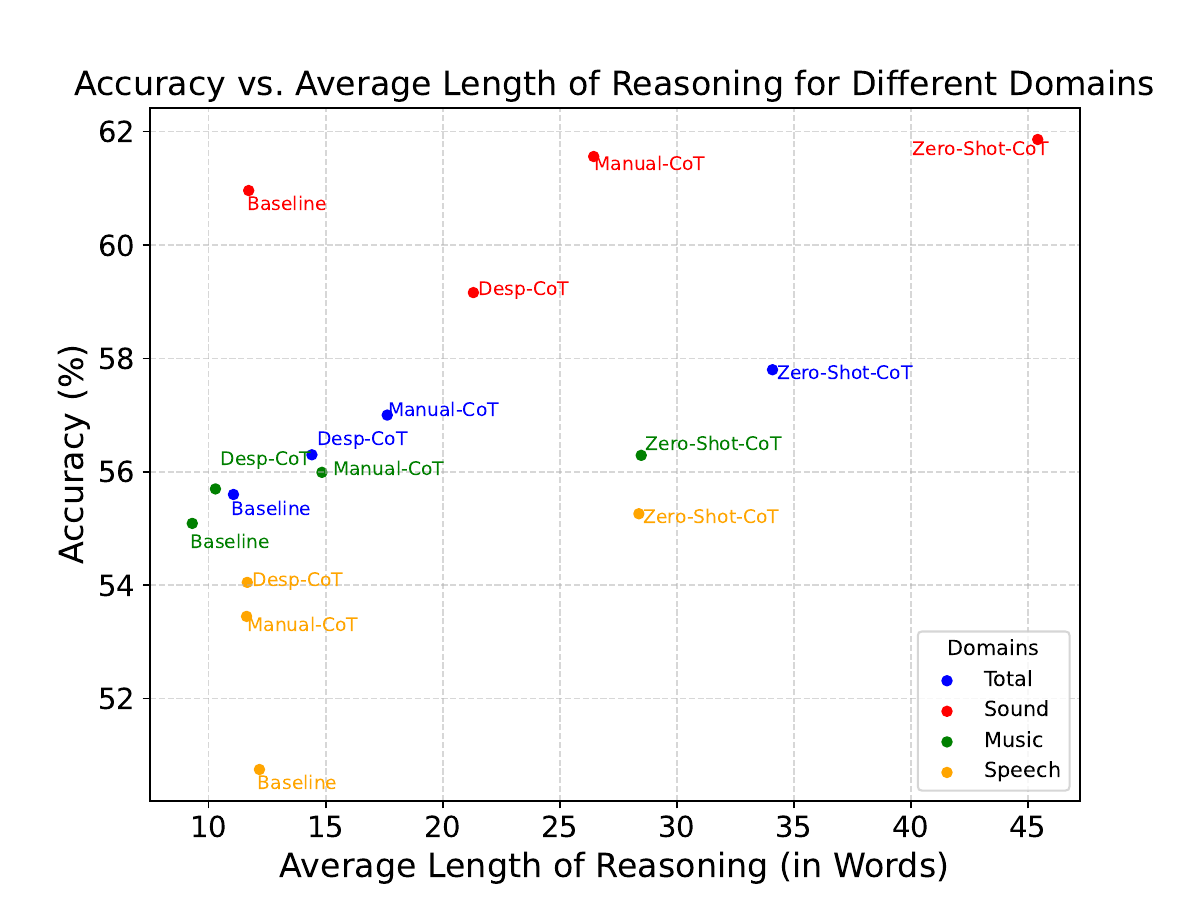}
  \caption{Comparison of accuracy and answer length (in words) for different methods across various domains. Each colored cluster represents one category, showing how the average reasoning length correlates with accuracy for each method.}
  \label{fig:length}
\end{figure}

We conducted an analysis of the relationship between reasoning length and accuracy across different CoT methods and observed a notable correlation between them. 
As illustrated in Figure~\ref{fig:length}, with the exception of the speech modality, and the Desp-CoT method in the sound modality, which appear as outliers, accuracy improves as reasoning paths become longer in the sound, music, and total categories. 
Notably, Zero-Shot-CoT exhibits the longest average reasoning length with approximately $35$ words, and achieves the highest performance. 
These findings indicate that current LALMs have the potential to enhance their instruction-following and reasoning capabilities by scaling the length of their inference processes. 

To evaluate the extent to which CoT methods can assist the model, we analyze the accuracy variation of different CoT approaches over the baseline across questions of varying difficulty levels: easy, medium, and hard, as shown in Figure~\ref{fig:difficulty}. 
Surprisingly, while CoT methods consistently enhance performance on easy and medium questions for both information extraction and reasoning tasks, they fail to improve accuracy on hard questions, and even result in performance degradation. 
This suggests that generating reasoning chains for more challenging questions may inadvertently confuse the model, highlighting the limitations of current LALMs and the need for further advancements in their foundational capabilities to handle more difficult tasks. 

\section{Conclusion \& Future Work}
In this work, we explore the Chain-of-Thought (CoT) reasoning methods in Large Audio-Language Models (LALMs), providing the first exploration in this domain. 
By leveraging the in-context reasoning capabilities of LALMs, we systematically evaluate various CoT approaches across diverse tasks, modalities, and difficulties. 
Our findings reveal the effectiveness of CoT methods on LALMs, the helpfulness of the Self-Consistency method for further performance improvement, the positiveness of reasoning length and performance correlation, as well as challenges with hard questions. 
While our study demonstrates the potential of CoT methods for LALMs, several challenges remain. Future research should focus on developing dynamic reasoning strategies to minimize the risk of confusion in challenging scenarios, and also enhancing LALM's foundational reasoning capabilities to handle complex reasoning tasks.

\newpage
\bibliographystyle{IEEEtran}
\bibliography{mybib}

\end{document}